# Non-uniform sampling, image recovery from sparse data and the discrete sampling theorem

Leonid P. Yaroslavsky, Gil Shabat, Benny G. Salomon, Ianir A. Ideses and Barak Fishbain[*]

*Department of Physical Electronics, Faculty of Engineering, Tel Aviv University,*

*Tel Aviv 69978, Israel*

[*]Corresponding author: barak@eng.tau.ac.il

## *Abstract*

In many applications sampled data are collected in irregular fashion or are partly lost or unavailable. In these cases it is required to convert irregularly sampled signals to regularly sampled ones or to restore missing data. In this paper, we address this problem in a framework of a discrete sampling theorem for "band-limited" discrete signals that have a limited number of non-zero transform coefficients in a certain transform domain. Conditions for the image unique recovery, from sparse samples, are formulated and then analyzed for various transforms. Applications are demonstrated on examples of image super-resolution and image reconstruction from sparse projections.

## 1 Introduction

Images and other signals are usually represented in computers in a form of their samples on a uniform sampling grid. However, in many applications sampled data are collected in irregular fashion and/or it may frequently happen that some samples of the regular sampling grid are lost or unavailable. In these cases it is required to convert irregularly sampled signals to regularly sampled ones or to restore missing data. Typical examples are filtering "salt &pepper"-type noise in images transmitted through communication channels with error detection coding, reconstruction of surface profiles in geophysics and in optical metrology, restoration of image sequences acquired in the presence of camera or object vibrations or through a turbulent medium and image super-resolution from multiple chaotically sampled frames, to name a few.

There are two approaches to treat this problem. One approach is empirical in nature and is based on simplistic numerical interpolation procedures such as, for instance, Shepard's interpolation by means of a weighted summation of known samples in close vicinity of sought samples with weights inversely proportional to the distance between them [1]. A review of these methods can be found in [2].

The second approach is based on generalizations of the classical Whittaker-Kotelnikov-Shannon sampling theory to non-uniform sampling. In this approach, it is assumed that the available signal samples are obtained from a continuous signal that belongs to a certain approximation subspace $M$ (e.g., sub-spaces of band-limited signals, splines subspaces, etc.) of the parent Hilbert space (usually, $L^2$ Hilbert space of finite energy functions) and it is required that the interpolation procedure has to determine a continuous signal that satisfies two constraints: 1) the interpolated signal has to belong to the subspace $M$ and 2) its available samples have to be preserved. Conditions for existence and uniqueness of the solution are dependent on the signal model (underlying approximation subspace) and the set of given samples. For the band-limited case,

Landau proved that a necessary and sufficient condition for the unique reconstruction of a continuous band-limited 1D signal with bandwidth $W$ from its irregularly spaced samples is that the density of its samples should exceed the Nyquist rate $1/W$ [3]. It is also shown that this condition is necessary for $D$-dimensional signals with band limited Fourier spectrum. These results have been generalized to other shift-invariant subspaces by Aldourbi and Grochenig [4]. A comprehensive presentation of this approach can be found in [5].

An attractive alternative approximation model is associated with spline subspaces [6]. However, due to their localized nature, their use for the recovery of large gaps in data is limited. A practical numerical algorithm for interpolation and approximation of 2D signals based on multilevel B-Splines is suggested by Wolberg *et al.* [7]. The algorithm approximates 2D functions from sparse data by an iterative procedure based on lattice control points. At each iteration, the values of available samples are preserved (if possible) or approximated. At the next iteration, a denser grid of control points is created to approximate the reconstruction error, and the process continues iteratively. Similar spline-based algorithm, which uses, for interpolation, non-uniform splines, was suggested by Margolis and Eldar [8].

All mentioned methods are theoretically oriented at the approximation of continuous signals, specified by their sparse samples. There are also publications that consider discrete models. However, those publications treat only various special cases. Fereira in [9] considers discrete signal recovery from sparse data in the assumption of signal band-limitation in the DFT domain. Hasan and Marvasti suggested a method for recovery discrete signals suffered from missing data during data transmission using error detecting coding. For signal recovery, they suggested using the DCT transform domain band limitation assumption [10]. In Ref. [11], the problem of non-uniform sampling in Fourier domain in multidimensional polar coordinates is addressed in connection with image reconstruction from projections. In yet another publication Averbuch and Zheludev discuss image reconstruction from projections with omissions using biorthogonal wavelets over-complete bases functions [12].

In this paper, we suggest a general framework for recovery of discrete signals, which originate from continuous signals, from incomplete sets of their samples. The base of this framework is the following assumptions:

- Continuous signals are represented in computers by their samples. In sampling a continuous signal, say $a(x)$, the physical coordinates of samples are known with certain accuracy. The ratio $N = X/\Delta x$ of the signal support interval $X$ and the sample position accuracy $\Delta x$ defines the signal regular uniform sampling grid with $N$ sampling positions. If all these $N$ samples were known, they would be sufficient for representing the continuous signal.

- Available are $K < N$ samples of this signal, taken at irregular positions of the signal regular sampling grid.

- The goal of the processing is generating, out of this incomplete set of $K$ samples, a complete set of $N$ signal samples in such a way as to secure the most accurate, in certain metrics, approximation of the discrete signal that corresponds to the signal, which would be obtained if the continuous signal it is intended to represent were densely sampled in all $N$ positions. For the certainty, we will use $L_2$ metrics.

The mathematical foundation of the framework is provided by the Discrete Sampling Theorem for "band-limited" discrete signals that have only few non-zero coefficients in their representation over certain orthogonal basis. This theorem is introduced in Sect. 2. The rest of the paper is as follows. In Sect. 3 we discuss the validity of the assumptions put in the base of the presented approach. In Sect. 4 we briefly describe a known iterative algorithm for signal recovery from sparse sampled data. In Sect. 5, the properties of certain transforms, which are specifically relevant for signal recovery from sparse data, are analyzed and experimental illustration of precise signal reconstruction from sparse data are provided. Finally, in Sect. 6 we discuss application issues and illustrate the discrete sampling theorem based methodology of discrete signal recovery on the examples of image super-resolution from multiple frames and image recovery from sparse projection data. Section 7 summarizes the paper.

## 2 Discrete Sampling Theorem

Let $\mathbf{A}_N$ be a vector of $N$ samples $\{a_k\}_{k=0,\ldots,N-1}$, which completely define a discrete signal, $\mathbf{\Phi}_N$ be an $N \times N$ orthogonal transform matrix

$$\mathbf{\Phi}_N = \{\varphi_r(k)\}_{r=0,1,\ldots,N-1} \tag{2-1}$$

and $\mathbf{\Gamma}_N$ be a vector of signal transform coefficients $\{\gamma_r\}_{r=0,\ldots,N-1}$ such that

$$\mathbf{A}_N = \mathbf{\Phi}_N \mathbf{\Gamma}_N = \left\{ \sum_{r=0}^{N-1} \gamma_r \varphi_r(k) \right\}_{k=0,1,\ldots N-1} \tag{2-2}$$

Assume now that available are only the $K < N$ samples $\{a_{\tilde{k}}\}_{\tilde{k} \in \tilde{\mathbf{K}}}$, where $\tilde{\mathbf{K}}$ is a $K$-size non-empty subset of indices $\{0,.1,\ldots,N-1\}$. These available $K$ signal samples define a system of $K$ equations:

$$\left\{ a_k = \sum_{r=0}^{N-1} \gamma_r \varphi_r(k) \right\}_{k \in \tilde{\mathbf{K}}} \tag{2-3}$$

for signal transform coefficients $\{\gamma_r\}$ of certain $K$ indices $r$.

Select now a subset $\tilde{\mathbf{R}}$ of $K$ transform coefficients indices $\{\tilde{r} \in \tilde{\mathbf{R}}\}$ and define a "**KofN**"-band-limited approximation $\hat{\mathbf{A}}_N^{BL}$ to the signal $\mathbf{A}_N$ as the

$$\hat{\mathbf{A}}_N^{BL} = \left\{ \hat{a}_k = \sum_{\tilde{r} \in \tilde{R}}^{1} \gamma_{\tilde{r}} \varphi_{\tilde{r}}(k) \right\} \tag{2-4}$$

Rewriteing this equation in a more general form:

$$\hat{\mathbf{A}}_N^{BL} = \left\{ \hat{a}_k = \sum_{r=0}^{N-1} \tilde{\gamma}_r \varphi_r(k) \right\} \tag{2-5}$$

And assuming that all transform coefficients with indices $r \notin \tilde{\mathbf{R}}$ are set to zero:

$$\tilde{\gamma}_r = \begin{cases} \gamma_r, & r \in \mathbf{R} \\ 0, & otherwise \end{cases} \tag{2-6}$$

Then the vector $\tilde{\mathbf{A}}_K$ of available signal samples $\{a_{\tilde{k}}\}$ can be expressed in terms of the basis functions $\{\varphi_r(k)\}$ of transform $\mathbf{\Phi}_N$ as:

$$\tilde{\mathbf{A}}_K = \mathbf{KofN}_\Phi \cdot \tilde{\mathbf{\Gamma}}_K = \left\{ \tilde{a}_{\tilde{k}} = \sum_{\tilde{r} \in \tilde{R}} \gamma_{\tilde{r}} \varphi_{\tilde{r}}(\tilde{k}) \right\} \tag{2-7}$$

where $K \times N$ sub-transform matrix $\mathbf{KofN}_\Phi$ is composed of samples $\varphi_{\tilde{r}}(\tilde{k})$ of the basis functions with indices $\{\tilde{r} \in \tilde{\mathbf{R}}\}$ for signal sample indices $\tilde{k} \in \tilde{\mathbf{K}}$, and $\tilde{\mathbf{\Gamma}}_K$ is a vector composed of the corresponding sub-set $\{\gamma_{\tilde{r}}\}$ of complete signal transform coefficients $\{\gamma_r\}$. This subset of the coefficients can be found as,

$$\tilde{\mathbf{\Gamma}}_K = \{\tilde{\gamma}_r\} = \mathbf{KofN}_\Phi^{-1} \cdot \tilde{\mathbf{A}}_K \tag{2-8}$$

provided matrix $\mathbf{KofN}_\Phi^{-1}$ inverse to the matrix $\mathbf{KofN}_\Phi$ exists, which, in general, is conditioned, for a specific transform, by positions $\tilde{k} \in \tilde{\mathbf{K}}$ of available signal samples and by the selection of the subset $\{\tilde{R}\}$ of transform basis functions.

By virtue of the Parceval's relationship for orthonormal transforms, the band-limited signal $\hat{\mathbf{A}}_N$ approximates complete signal $\mathbf{A}_N$ with mean squared error:

$$MSE = \left\|A_N - \hat{A}_N\right\| = \sum_{k=0}^{N-1}\left|a_k - \hat{a}_k\right|^2 = \sum_{r \notin R}\left|\gamma_r\right|^2 \qquad (2\text{-}9)$$

This error can be minimized by an appropriate selection of the $K$ basis functions of the sub-transform $\mathbf{KofN}_\Phi$. In order to do so, one must know the energy compaction ordering of basis functions of the transform $\mathbf{\Phi}_N$. If, in addition, one knows, for a class of signals, a transform that features the best energy compaction in the smallest number of transform coefficients, one can, by selection of this transform, secure the best band-limited approximation of the signal $\{a_k\}$ for the given subset $\{\tilde{a}_k\}$ of its samples.

In this way we arrive at the following Discrete Sampling Theorem that can be formulated in these two statements:

*<u>Statement 1.</u> For any discrete signal of $N$ samples defined by its $K \leq N$ sparse and not necessarily regularly arranged samples, its band-limited, in terms of certain transform $\mathbf{\Phi}_N$, approximation can be obtained with mean square error defined by Eq. (2-9). The approximation error can be minimized by using a transform with the best energy compaction property.*

*<u>Statement 2.</u> Any signal of $N$ samples that is known to have only $K \leq N$ non-zero transform coefficients for certain transform $\mathbf{\Phi}_N$ ($\mathbf{\Phi}_N$ - transform "band-limited" signal)*

*can be fully recovered from exactly **K** of its samples provided the positions of the sample secure the existence of the matrix* $\mathbf{KofN}_\Phi^{-1}$ *inverse to the sub-transform matrix* $\mathbf{KofN}_\Phi$ *that corresponds to the band-limitation.*

## 3  Validity of the assumptions

The applicability of the above results depends on the validity of the assumption that "band-limited", in certain basis, approximation of signals is an acceptable solution in image recovery. We believe that this assumption is validated by a consensus in signal processing and image processing community regarding signal compression, where such transforms as DCT and certain wavelets are known for their very good energy compaction properties for wide variety of signals in image and audio processing and are successfully used for compression by means of replacement of signals by their "band-limited" approximations. Recent advances in "compressive sensing" [13] also are based on signal "band-limitedness" assumption. Haar transform and Walsh transform were found to have good energy compaction properties for bi-level images such as drawings and documents. An important application, in which the assumption of image bound-limitedness is supported by the physical reality, is computed tomography, where slice projections can very frequently be regarded as band-limited, in inverse Radon transform domain, signals because outer parts of slices are usually known to be empty.

# 4 Iterative algorithm for signal recovery from sparse non-uniformly sampled data

Implementation of signal recovery from sparse non-uniformly sampled data according to Eq. (2-8) requires matrix inversion, which is, generally, a very computationally demanding procedure. In applications, one can always be satisfied with signal reconstruction with certain limited accuracy and apply for the reconstruction a simple iterative reconstruction procedure of the Gershberg-Papoulis [14] type shown in flow diagram of Fig. 1. We used this algorithm in the experiments reported in this paper. A review of other iterative and non-iterative algorithmic options one can find in [9].

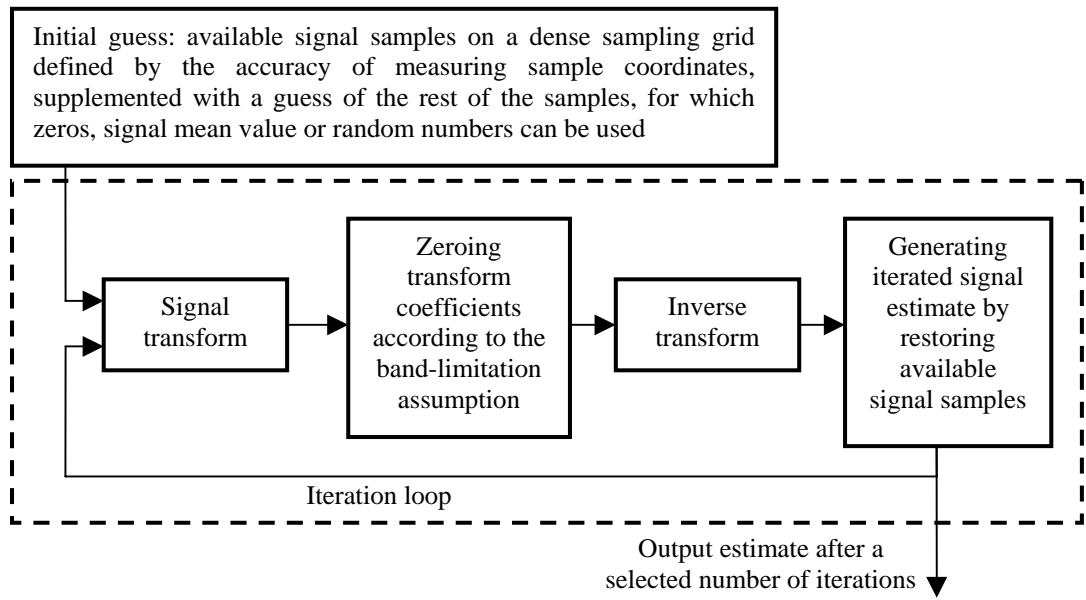

Fig. 1 - Flow diagram of the iterative signal recovery procedure

# 5 Analysis of transforms

## 5.1 Discrete Fourier Transform

Consider the $\mathbf{K}of\mathbf{N}_{DFT}^{LP}$-trimmed $DFT_N$ matrix:

$$\mathbf{K}of\mathbf{N}_{DFT}^{LP} = \left\{ \exp\left( i2\pi \frac{\tilde{k}\tilde{r}_{LP}}{N} \right) \right\} \tag{5-1}$$

that corresponds to DFT $\mathbf{K}of\mathbf{N}$-low-pass band-limited signal. Due to complex conjugate symmetry of DFT or real signals, $K$ has to be an odd number, and the set of frequency domain indices of $\mathbf{K}of\mathbf{N}_{DFT}$ low-pass band-limited signals in Eq. (5-1) is defined as:

$$\tilde{r}_{LP} \in \tilde{R}_{LP} = \{0,1,...,(K-1)/2, N-(K-1)/2,...,N-1\} \tag{5-2}$$

For such a case, the following theorems hold:

Theorem 1.

*$\mathbf{K}of\mathbf{N}$-low-pass DFT band-limited signals of $N$ samples with only $K$ nonzero low frequency DFT coefficients can be precisely recovered from exactly $K$ of their samples taken in arbitrary positions.*

Proof.

As it follows from Eqs. (2-3)-(2-8), the theorem is proven if matrix $\mathbf{K}of\mathbf{N}_{DFT}^{LP}$ is invertible. A matrix is invertible if its determinant is nonzero. In order to check whether determinant of the matrix $\mathbf{K}of\mathbf{N}_{DFT}$ is non-zero, permute the order of columns of the matrix as following:

$$\tilde{\tilde{r}} \in \tilde{\tilde{R}} = \{N-(K-1)/2,...,N-1,0,1,...,(K-1)/2\} \tag{5-3}$$

and obtain matrix

$$\mathbf{K}of\mathbf{N}_{DFT}^{DFTsh} = \left\{\exp\left[i2\pi\frac{\tilde{k}\tilde{\tilde{r}}}{N}\right]\right\} =$$
$$= \left\{\exp\left[i2\pi\frac{N-(K-1)/2}{N}\tilde{k}\right]\delta\left(\tilde{k}-\tilde{\tilde{r}}\right)\right\}\left\{\exp\left[i2\pi\frac{\tilde{k}\tilde{\tilde{r}}}{N}\right]\right\} \quad (5\text{-}4)$$

where

$$\tilde{\tilde{r}} \in \tilde{\tilde{R}} = \{[0,...,K-1]\} \quad (5\text{-}5)$$

The first matrix in this product of matrices is a diagonal matrix, which is obviously invertible. The second one is a version of Vandermonde matrices, which are also known to have non-zero determinant if, like in our case, its ratios for each row are distinct [15].

As permutation of the matrix columns does not change the absolute value of its determinant, Eq. (5-4) implies that determinant of $\mathbf{K}of\mathbf{N}$-trimmed $DFT_N$ matrix of Eq. (5-1) is also non-zero for arbitrary set $\tilde{K} = \{\tilde{k}\}$ of positions of $K$ available signal samples.

One can easily see that for DFT $\mathbf{K}of\mathbf{N}$-high-pass band-limited signals, for which

$$\mathbf{K}of\mathbf{N}_{DFT}^{HP} = \left\{\exp\left(i2\pi\frac{\tilde{k}\tilde{r}_{HP}}{N}\right)\right\} \quad (5\text{-}6)$$

where

$$\tilde{r}_{HP} \in \tilde{R}_{HP} = \{[(N-K+1)/2, (N-K+3)/2,..., (N+K-1)/2]\} \quad (5\text{-}7)$$

a similar theorem holds

Theorem 2.

*$\mathbf{K}of\mathbf{N}$-high-pass DFT band-limited signals of N samples with only K nonzero high frequency DFT coefficients can be precisely recovered from exactly K of their arbitrarily taken samples.*

Note that, due to the complex conjugate symmetry of DFT of real signals, *K* in this case has to be odd whatever *N* is.

Obviously, above Theorems 1 and 2 can be extended to a more general case of signal DFT band limitation, when indices $\{\tilde{r}\}$ of nonzero DFT spectral coefficients form arithmetic progressions with common difference other than one such as, for instance,

$$\tilde{r}_{mLP} \in \tilde{R}_{mLP} =$$
$$= \{0, m, ..., m(K-1)/2, N - m(K-1)/2, ..., N - m(K-1)/2 + (K+1)/ \quad (5\text{-}8)$$

## 5.2 Discrete Cosine Transform (DCT)

*N*-point Discrete Cosine Transform of a signal is equivalent to 2*N*-point Shifted Discrete Fourier Transform (SDFT) with shift parameters (1/2,0) of 2*N*- sample signal obtained from the initial one by its mirror reflection [16]. **K***of***N** -trimmed matrix of SDFT(1/2,0)

$$\mathbf{K}of\mathbf{N}_{SDFT} = \left\{ \exp\left( i2\pi \frac{(\tilde{k}+1/2)\tilde{r}}{2N} \right) \right\} \quad (5\text{-}1)$$

can be represented as a product

$$\mathbf{K}of\mathbf{N}_{SDFT} = \left\{ \exp\left( i2\pi \frac{\tilde{k}\tilde{r}}{2N} \right) \right\} \left\{ \exp\left( i\pi \frac{\tilde{r}}{2N} \right) \delta(k-r) \right\} =$$
$$= \mathbf{K}of\mathbf{N}_{DFT} \left\{ \exp\left( i\pi \frac{\tilde{r}}{2N} \right) \delta(k-r) \right\} \quad (5\text{-}2)$$

of a **2*N*** -point DFT matrix and a diagonal matrix $\left\{ \exp\left( i\pi \frac{\tilde{r}}{2N} \right) \delta(k-r) \right\}$. The latter one is invertible and the invertibility of **K***of***N** -trimmed $DFT_{2N}$ matrix $\mathbf{K}of\mathbf{N}_{DFT}$ can be proved, for above described band-limitations, as it was done above for the DFT case. Therefore, for DCT theorems similar to those for DFT hold.

These theorems hold also for 2D DFT and DCT transforms provided band-limitation conditions are separable. The case of non-separable band-limitation requires further study. In the discussion of experiments that follows we will compare separable and non-separable band-limitation in DCT domain. Note that working in DFT or DCT domain results, in the case of low-pass band-limitation, in signal discrete sinc-interpolation [17].

We illustrate the above reasoning by some simulation examples. The plots in Fig. 2 illustrate exact reconstruction of a DFT-"band-limited" signal (plotted in red) for two cases, when all available signal samples form a compact group (Fig. 2, left top) and when they are randomly placed within signal support (Fig. 2, left bottom). The right hand side of Fig. 2, illustrates restoration of the same signal with randomly placed samples by means of the iterative algorithm. Note that the speed of convergence of the iterative algorithm heavily depends on the realization of sample positions and, for some realizations of sample positions might be very slow.

Fig. 3 and Fig. 4 illustrate precise restoration from sparse data of images band-limited in DCT domain by a square (separable band-limitation) and by $90^o$ circle sector (a pie piece, inseparable band-limitation). In these experiments, image restoration using multilevel B-spline interpolation algorithm was used as a benchmark [7] [1]. The image presented in Fig. 3, is a **64×64** pixel test image low-pass band-limited in DCT domain by a **14×14** sample square (Fig. 3, b). It has only **14×14 = 196** nonzero DCT spectral components out of the **64×64** signal samples. This image was sampled at **196** "random" positions obtained from a standard Matlab pseudo-random number generator. One can see from the figure, that iterative algorithm provides quite accurate restoration of the initial image, though precise restoration may require quite large number of iterations. An important peculiarity in 2D case the convergence of iteration is very non-uniform within the image. Usually, the restoration error is rapidly becoming very small almost everywhere in the image, and only in some parts, where sample density happens to be low, the restoration errors remain to be substantial and converge to zero quite slowly.

---

[1] For the implementation of the multilevel B-Splines algorithm, a code kindly provided by Prof. Wolberg was used.

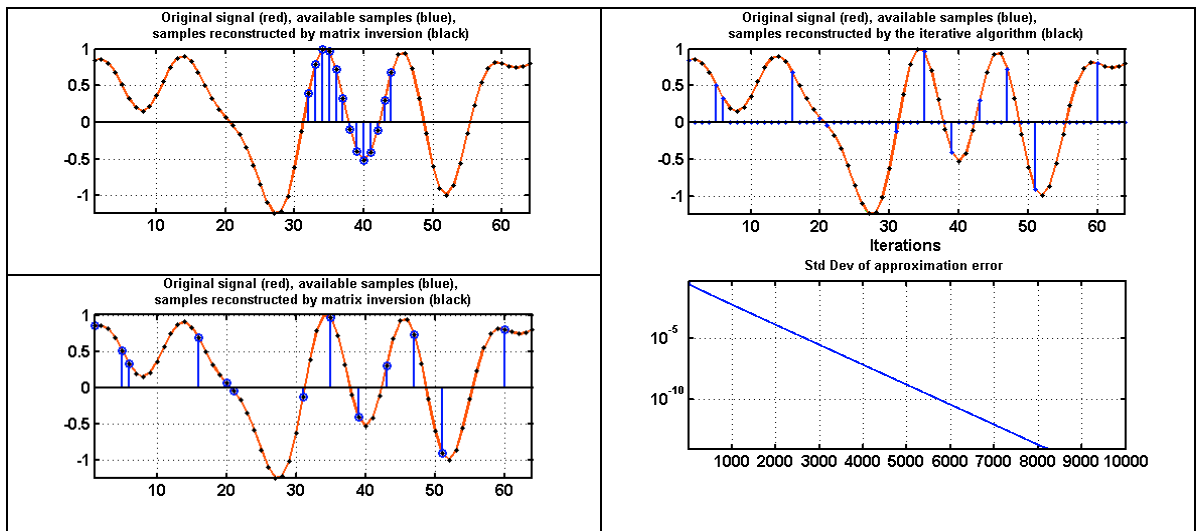

Fig. 2 - Restoration of a DFT low pass band-limited signal by matrix inversion for the cases of random (a), upper) and compactly placed signal samples (a), bottom) and by the iterative algorithm (b). Bottom right plot shows standard deviation of signal restoration error as a function of the number of iterations. The experiment was conducted for test signal length 64 samples; bandwidth 13 frequency samples (~1/5 of the signal base band)

Image band limitation by a square is separable and, as was shown earlier, it does not impose any limitations on the positions of sparse samples. It is, however, not isotropic. In the case of isotropic band limitation in DCT domain by a circle sector (a pie piece), the situation is quite different. Experiments show that the speed of convergence of the iterative algorithm dramatically drops in this case. Hundreds of thousands of iterations are needed to make standard deviation of the restoration error lower than 0.1, though again, restoration error remains to be substantial only in limited areas of the image. B-spline interpolation error, in this case, is also high, though it is uniform over the image. The convergence speed of the iterative algorithm in the case of isotropic circle sector band limitation can be substantially improved if the number of available image samples exceeds the number of non-zero DCT spectral coefficients, which are redundant from the point of view of the Discrete Sampling Theorem. This is illustrated in Fig. 4. The image presented in Fig. 4 is a $64 \times 64$ pixel test image, which is low-pass band-limited in DCT domain by a circle sector. It has $196$ nonzero DCT spectral components, out of $64 \times 64$ signal's samples, all located within a circle sector shown in white in Fig. 4, b). In distinction to the image of Fig. 3, this one was sampled at $248$ "random" positions. The redundancy $248/196 = 1.27$ in the number of samples with respect to the number of non-zero spectral coefficients is approximately equal to the ratio of the area of a square to the area of the circle sector inscribed into this square. As one can see from Fig. 4, f), with such a redundancy, iterative restoration converges much faster, though overall restoration error even after 100,000 iterations remains higher than that for the separable band limitation by a square illustrated in Fig. 3. The same holds for B-spline interpolation restoration, shown in Fig. 4, d). Once again, one can see that the convergence of the iterative algorithm is substantially non-uniform over the image and relatively large restoration error occurs only in a small area of the image where the density of available samples happens to be low.

In some applications, there is a natural and substantial redundancy in the number of available image samples with respect to its bandwidth. One of such cases is illustrated in Fig. 5, where an example of image restoration from its level lines is given. $256 \times 256$ pixel image shown in the figure is band limited in DCT domain by a circle sector and contains 302 non-zero spectral coefficients. The image was sampled in 6644 samples on a set of its level lines (8 levels), which resulted in 22-fold redundancy with respect to the image spectrum. As one can see from the figure, such a redundancy accelerated the convergence of the iterative algorithm very substantially and enabled, after a few tens of iterations, restoration, which is much superior with respect to that provided by the B-spline interpolation.

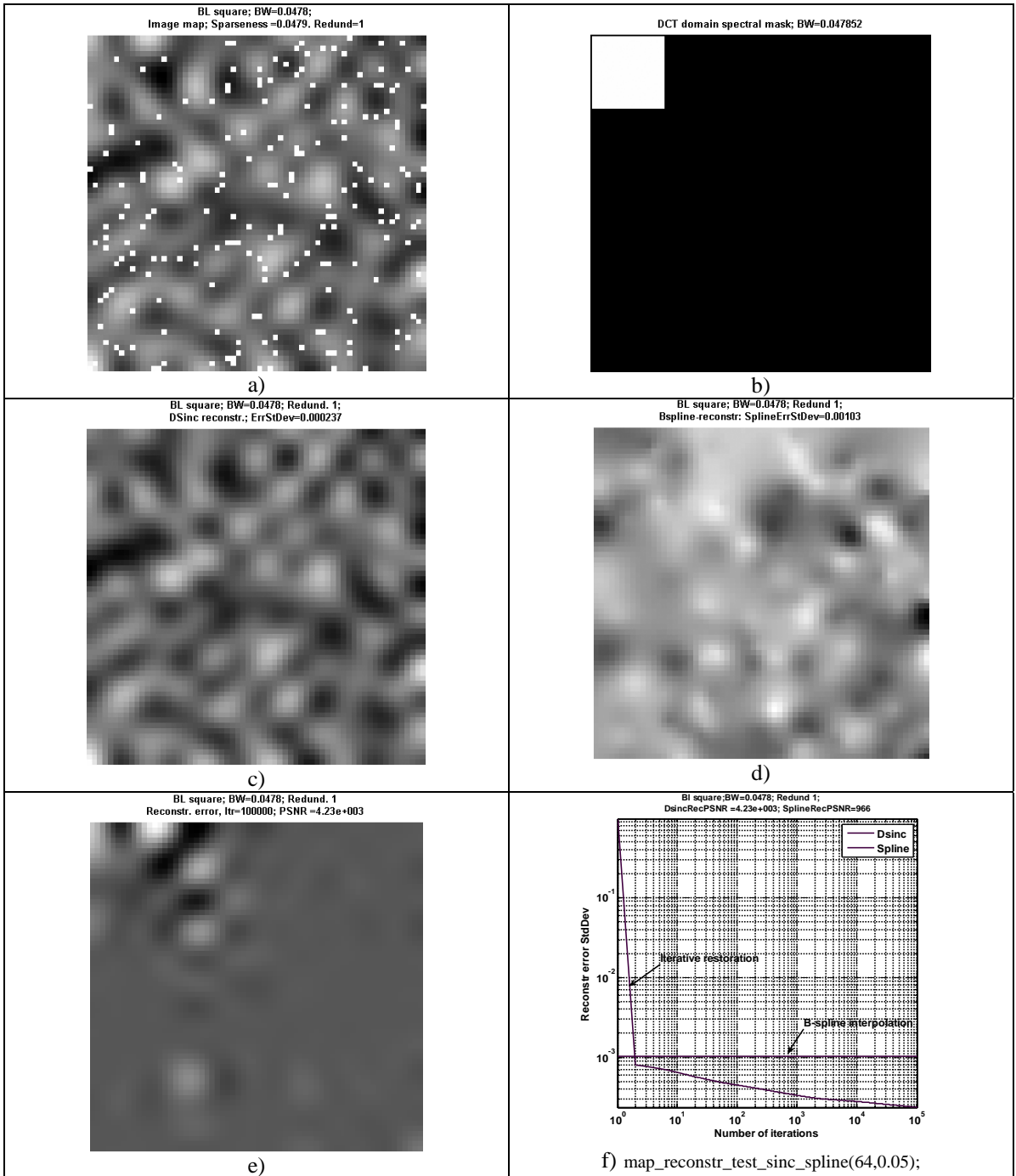

Fig. 3 - Recovery of an image band limited in DCT domain by a square: a) – initial image with 3136 "randomly" place samples (shown by white dots); b) – the shape of the image spectrum in DCT domain; c) –image restored by the iterative algorithm after 100000 iterations with restoration PSNR (peak signal-to-error standard deviation) 4230; d) image restored by B-spline interpolation with restoration PSNR 966; e) iterative algorithm restoration error (white – large errors; black – small errors); f) –restoration error standard deviation versus the number of iterations for the iterative algorith and that for the B-spline interpolation

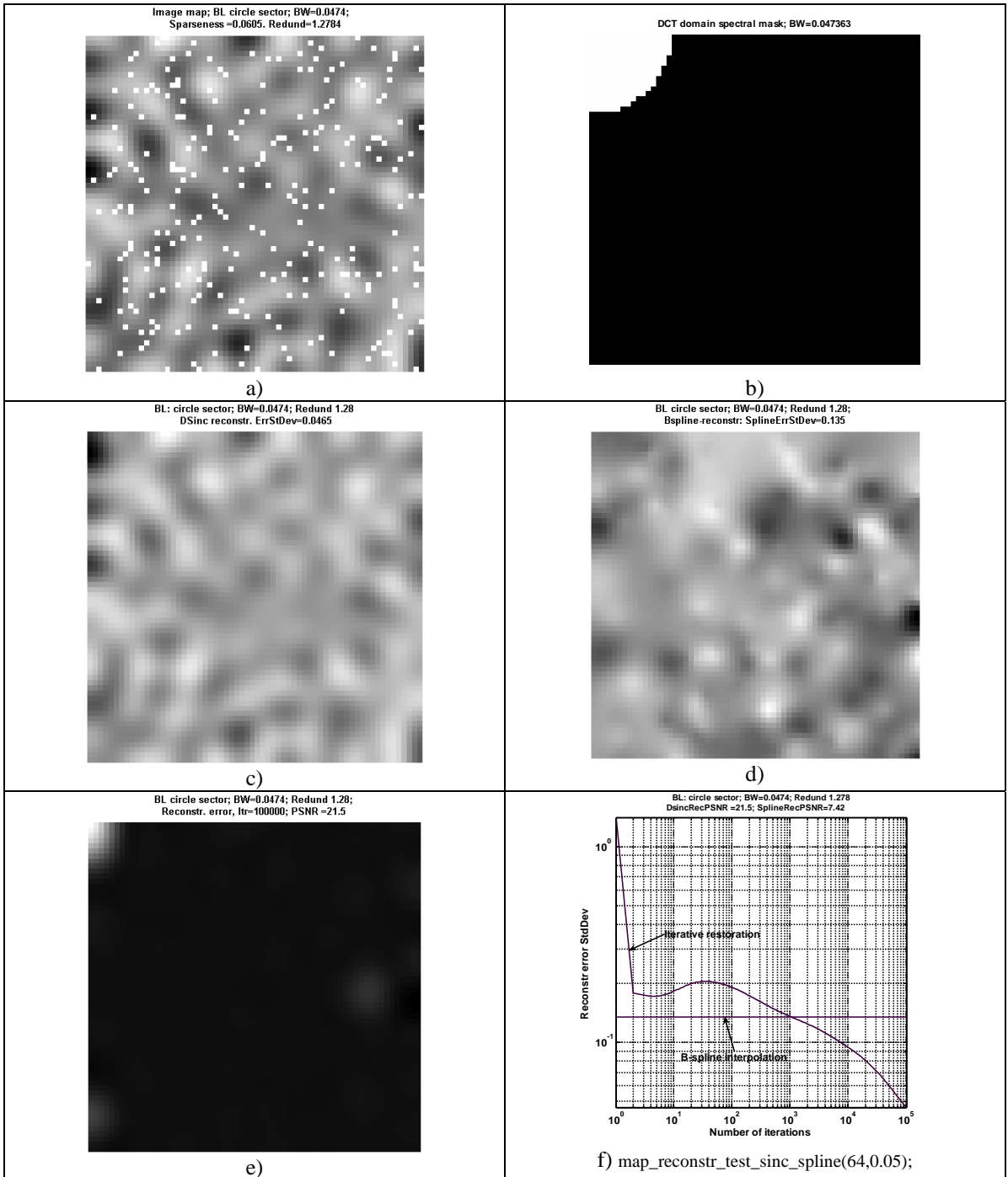

Fig. 4 - Recovery of an image band limited in DCT domain by a circle sector: a) – initial image with 3964 "randomly" place samples (shown by white dots); b) – the shape of the image spectrum in DCT domain; c) –image restored by the iterative algorithm after 100000 iterations with restoration PSNR (peak signal-to-error standard deviation) 21.5; d) image restored by B-spline interpolation with restoration PSNR 7.42; e) iterative algorithm restoration error (white – large errors; black – small errors); f) –the restoration error standard deviation versus the number of iterations of the iterative algorithm for the iterative algorithm and that for the B-spline interpolation

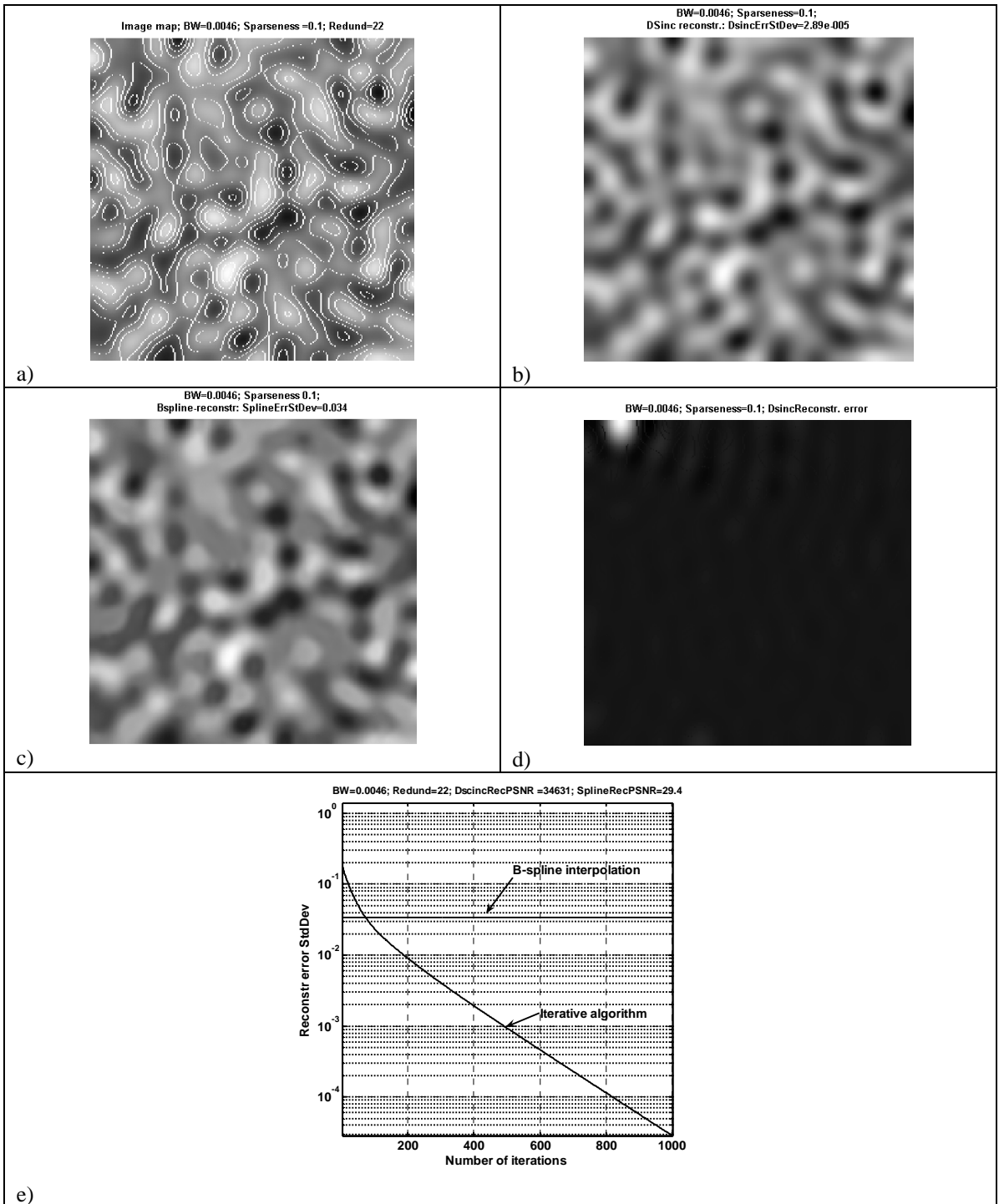

Fig. 5 - Recovery of an image band limited in DCT domain by a circle sector from its level lines: a) – initial image with level lines (shown by white dots); b) –image restored by the iterative algorithm after 1000 iterations with restoration PSNR $3.5 \times 10^4$ (note that the restoration error is concentrated in a small area of the image); c) image restored by B-spline interpolation with restoration PSNR 29.4; d) iterative algorithm restoration error (white – large errors; black – small errors); e) –the restoration error standard deviation versus the number of iterations of the iterative algorithm for the iterative algorithm and that for the B-spline interpolation

## 5.3 Wavelets and Other Bases

The main peculiarity of wavelet bases is that their basis functions are most naturally ordered in terms of two components: scale and position within the scale. Scale index is analogous to the frequency index for DFT. Position index tells only of the shift of the same basis function within the signal on each scale. Therefore band-limitation for DFT translates to scale limitation for wavelets. Limitation in terms of position is trivial: it simply means that some parts of the signal are not relevant. Commonly, discrete wavelets are designed for signals whose length is an integer power of 2 ($N = 2^n$). For such signals, there are $s \leq n$ scales and possible "band-limitations".

The simplest special case of wavelet bases is Haar basis. Signals with $N = 2^n$ samples and only with $K$ lower index non-zero Haar transform (the transform coefficients $\{K,...,N-1\}$ are zero) are ($\tilde{s} = (\lfloor \log_2(K-1) \rfloor + 1)$) - "band-limited", where $\lfloor x \rfloor$ is an integer part of $x$. Such signals are piecewise constant within intervals between zero-crossings. The shortest intervals of the signal constancy have $2^{n-\tilde{s}}$ samples. As one can see from Fig. 6, a), for any two samples that are located on the same interval, all Haar basis function on this and lower scales have the same value. Therefore, having more than one sample per constant interval will not change the rank of the matrix **KofN**. The condition for perfect reconstruction is, therefore, to have at least one sample on each of those intervals.

For other wavelets as well as for other bases general necessary, sufficient and easily verified condition for the invertibility of **KofN**-trimmed transform sub-matrix is not known for the present authors. Standard linear algebra procedures for determining matrix rank, can be used for testing invertibility of the matrix.

For Walsh basis functions, the index corresponds to the "sequency", or to the number of zero crossings of the basis function. The sequency carries a certain analogy to the signal frequency. Basis functions ordering according to their sequency, which is characteristic for Walsh transform, preserves, for many real signals, the property of decaying transform coefficients' energy with their index. Therefore, for Walsh transform the notion of low-pass band-limited signal approximation, similar to the one described in Sect. 5.1, for DFT, can be used. On the other hand, as one can see from Fig. 6, b), Walsh basis functions, similarly to Haar basis function, can be characterized by the scale index, which specifies the shortest interval of signal constancy. Signals with $N = 2^n$ samples and band-limitation of $K$ Walsh transform coefficients have shortest intervals of signal constancy of $2^{n-\tilde{s}}$ samples, where $\tilde{s} = (\lfloor \log_2(K-1) \rfloor + 1)$. A necessary condition for perfect reconstruction is to have $K$ signal samples taken on different intervals. Unlike the Haar transform case, not all the intervals are needed to be sampled, but only $K$ intervals out off the total number of intervals. For a special case of $K$ equal to a power of 2, there are $K$ intervals, each of which has to be sampled to secure perfect reconstruction, This is the case, when the reconstruction condition for Walsh Transform is identical to that for Haar transform.

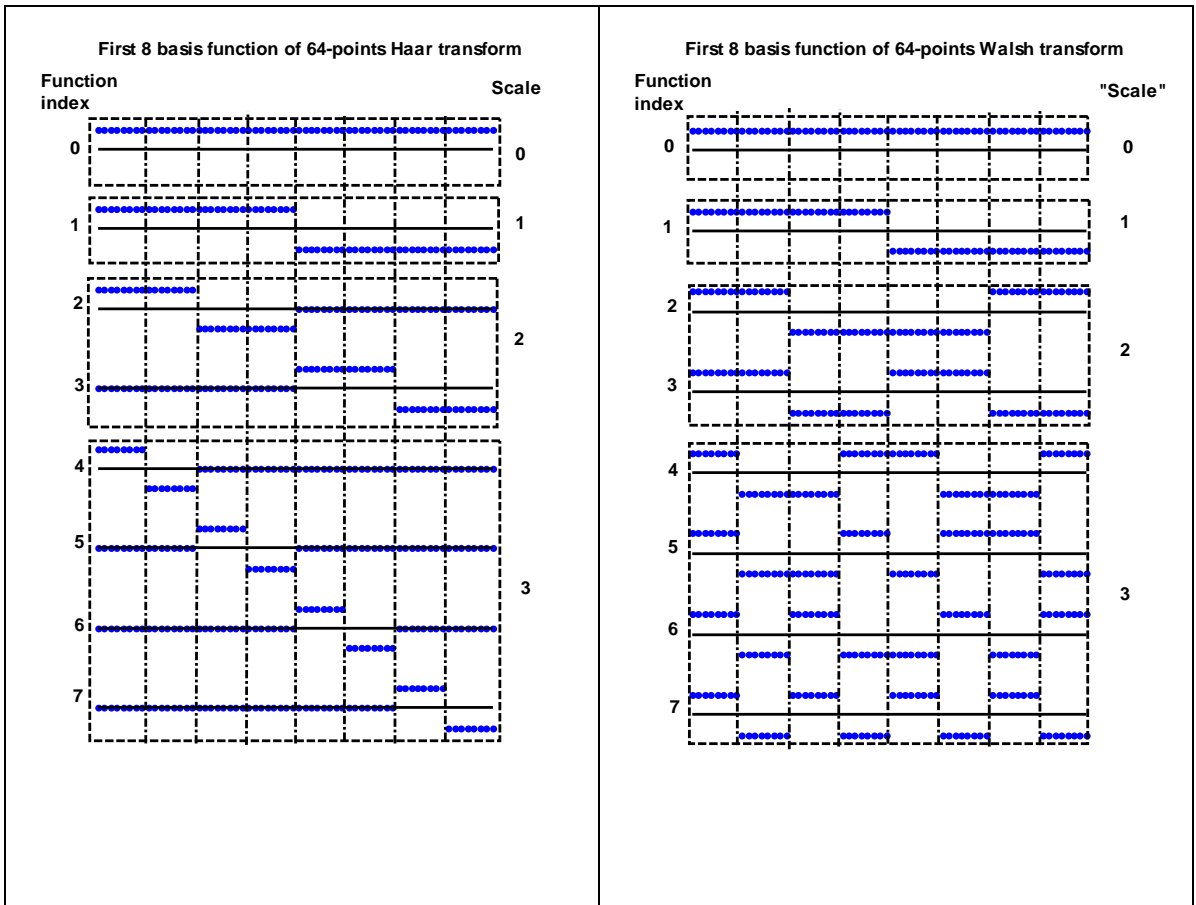

Fig. 6 - First 8 basis functions of 64 point Haar (a) and Walsh (b) transforms. Intervals of function constancy are outlined by dash-dot lines. Functions that belong to the same scale are outlined by dashed boxes.

Fig. 7 illustrates the case of recovery of an image "band limited" in the Haar transform domain. Two examples are shown: arrangement of sparse samples, for which signal recovery is possible (a) and that for which signal is not recoverable (b). Note that when the Haar reconstruction is possible, it is reduced to the trivial nearest neighbor interpolation.

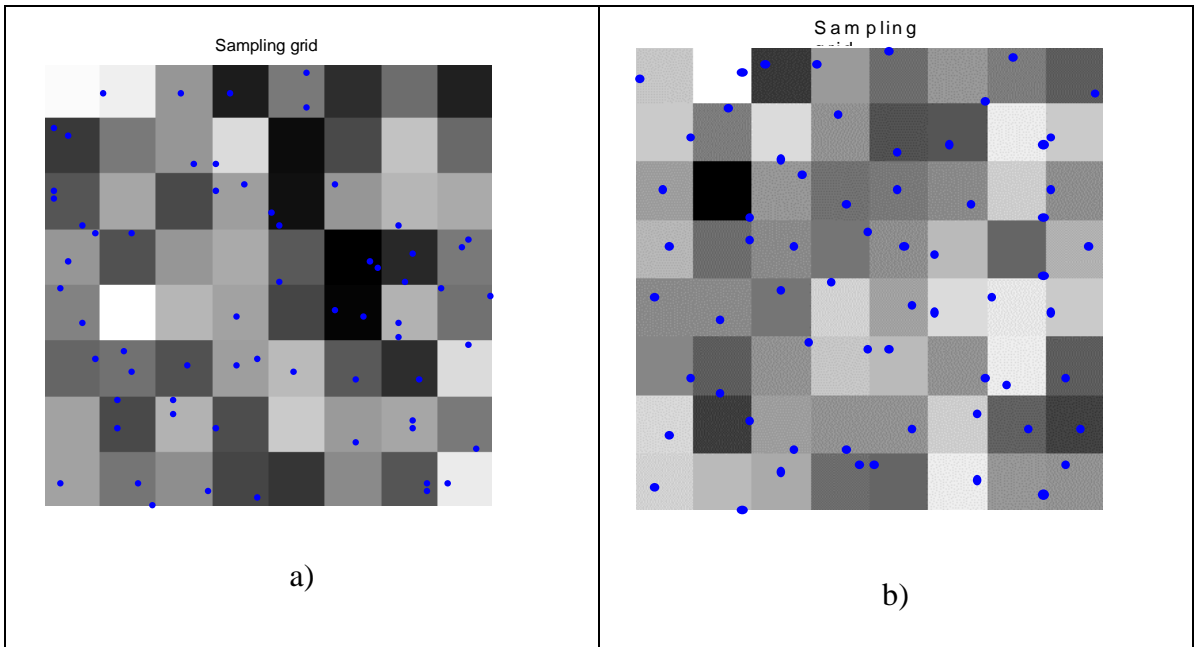

Fig. 7 - Two cases of sparse sampling of an image band-limited in Haar Transform: a) not recoverable case; b) recoverable case (sample points are marked with dots). Image size 64x64 pixels; band-limitation 8x8 (scale 3)

An example of perfect reconstruction of Walsh transform domain "band-limited" signal of $N$=512 and band limitation $K$=5 is illustrated in Fig. 8. In this example, the resulted $KofN^{Walsh}$ matrix is:

$$KofN^{Walsh}\Big|_{K=5} = \begin{bmatrix} 1 & -1 & 1 & -1 & -1 \\ 1 & -1 & -1 & 1 & 1 \\ 1 & 1 & 1 & 1 & 1 \\ 1 & 1 & -1 & -1 & -1 \\ 1 & 1 & 1 & 1 & -1 \end{bmatrix} \quad (5\text{-}1)$$

and its rank equals to 5. One should note that, in this particular example, perfect reconstruction in the Haar transform domain is not possible since one of the shortest intervals of the signal constancy contains no samples.

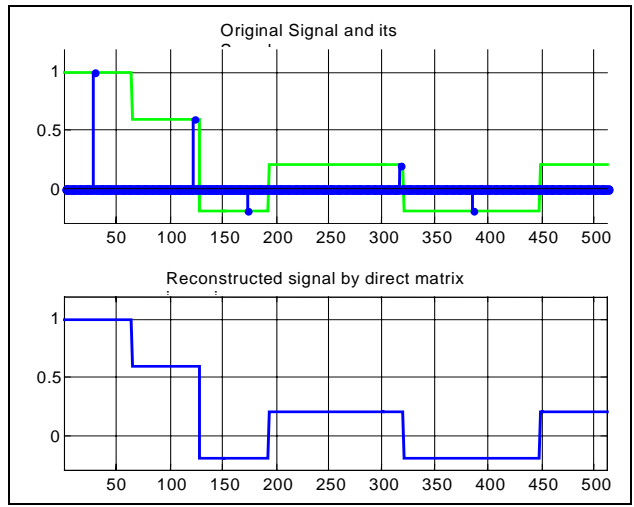

Fig. 8 - Example for perfect reconstruction on Walsh domain

# 6 Application examples

## *6.1 Image super-resolution from multiple differently sampled video frames.*

One of the potential applications of the above signal recovery technique is image super-resolution from multiple video frames with chaotic pixel displacements due to atmospheric turbulence, camera instability or similar random factors [18]. By means of elastic registration of sequence of frames of the same scene, one can determine, for each image frame and with sub-pixel accuracy, pixel displacements caused by random acquisition factors. Using these data, a synthetic fused image can be generated by placing pixels from all available video frames in their proper positions on the correspondingly denser sampling grid according to their found displacements. In this process, some pixel positions on the denser sampling grid will **remain** unoccupied, especially when limited number of image frames is fused. These missing pixels can then be restored using the above-described iterative band-limited interpolation algorithm.

In the implementation of the algorithm, the denser sampling grid of the fused image is formed accordingly to the sub-pixel accuracy, with which positions of pixel are measured in the sequence of turbulent frames. In our experiment, the size of the fused image sampling grid was 8 times that of initial frames. The bandwidth limitation of the super-resolved image depends on the spread of image samples acquired in the process of fusion and the number of frames used for fusion. In our experiments, we set final size of the fused image sampling grid to be twice that of original frames. The simulation result of iterative recovery of unavailable image samples is presented in Fig. 9, which shows one of low resolution turbulent frames (a), image fused from 50 frames (b) and a result of iterative interpolation (c) achieved after 50 iterations. It clearly demonstrates that a substantial improvement of image resolution and quality is possible.

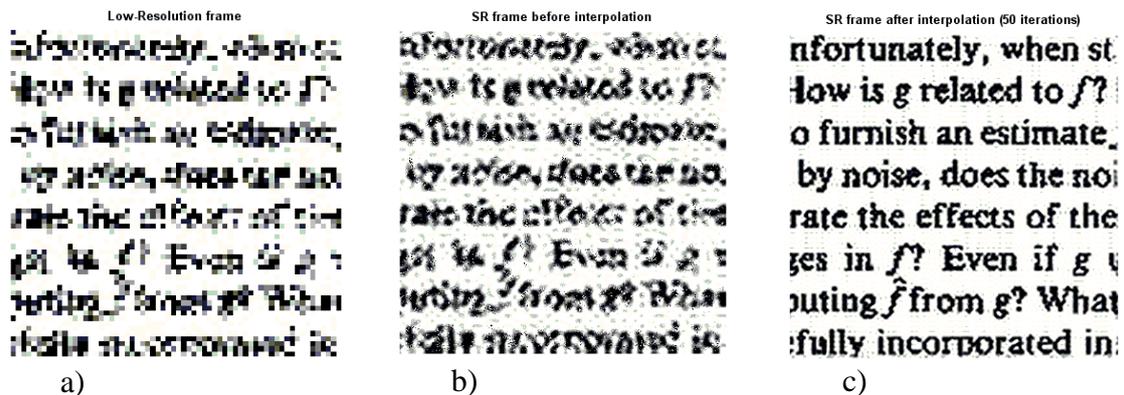

Fig. 9 - Iterative image interpolation in the super-resolution process: a) – a low resolution frame; b) image fused by elastic image registration from 50 frames; c) – a result of iterative interpolation of image b) after 50 iterations.

## 6.2 Image reconstruction from sparse projections in computed tomography

A straightforward application the discussed sparse data recovery algorithm can found in tomography imaging, where it frequently happens that a substantial part of slices, which surrounds the body slice, is known to be an empty field. This means that slice projections (sinograms) are Radon transform "band-limited" functions. Therefore whatever number of projections is available, a certain number of additional projections, commensurable, according to the discrete sampling theorem, with the size of the slice empty zone, can be obtained and the corresponding resolution increase in the reconstructed images can be achieved using the described iterative band-limited reconstruction algorithm. Another option is recovery of projection data that might be missing due to sensor faults or to other reasons.

In order to demonstrate the applicability of the discrete sampling theorem for image recovery from sparse projections, one needs a discrete Radon transform and its algebraically exact inverse. While the theory defines the continuous Radon integral transform and its inverse, the discrete equivalent is not a trivial problem. In our experiments we used a stable forward and inverse Radon transform described in [19] and the code found in [20]. The applicability the discrete Radon transform within the suggested framework is illustrated in Fig. 10. By simple segmentation of the initial image shown in Fig. 10, a) it was found that the outer 55% of the image area is empty. Then the same percentage of projection samples selected randomly using the Matlab random number generator were zeroed after which the iterative reconstruction algorithm was run. The results shown in Fig. 10, (c) through (f), show that while direct image reconstruction with missing samples completely fails (Fig. 10, c), virtually perfect recovery of missing 55% samples of sinograms is possible with the iterative reconstruction algorithm after several hundreds of iterations.

Fig. 11 illustrates that recovery of completely missing projections is also possible. Every second of projections of image shown in Fig. 10, a) was removed and then all initial projections were recovered by the iterative algorithm that made use of the fact that the outer 55% part of the image area is known to be empty. In this case the standard deviation of the reconstruction error is not as low as in the previous case, which, perhaps, can be attributed to not full reversibility of the truncated Radon Transforms. However, the achieved low reconstruction error of about $10^{-3}$ allows to suggest that for such cases, when half or bigger part of the image area is known to be empty, one can achieve image reconstruction with super-resolution that corresponds to double number of available image projections.

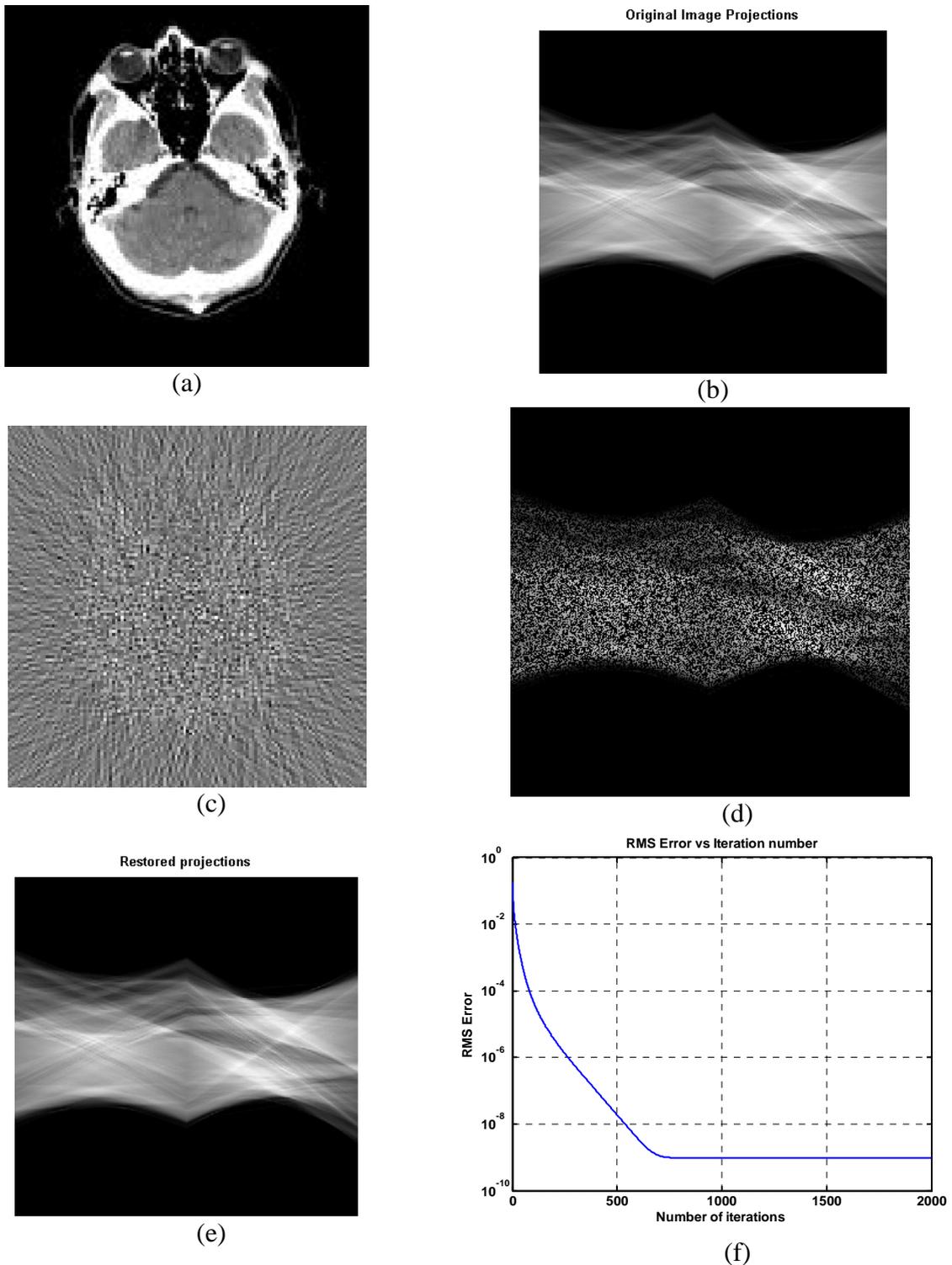

Fig. 10 - Recovery of missing samples of a sinogram: (a), (b) original image and its Radon transform (sinogram), (c) image reconstructed from the sinogram (d) corrupted by the loss of 55% of its randomly selected samples; e) a sinogram recovered from (d) using the iterative band-limited interpolation algorithm and (f) a plot of standard deviation of slice reconstruction error as a function of the iteration number.

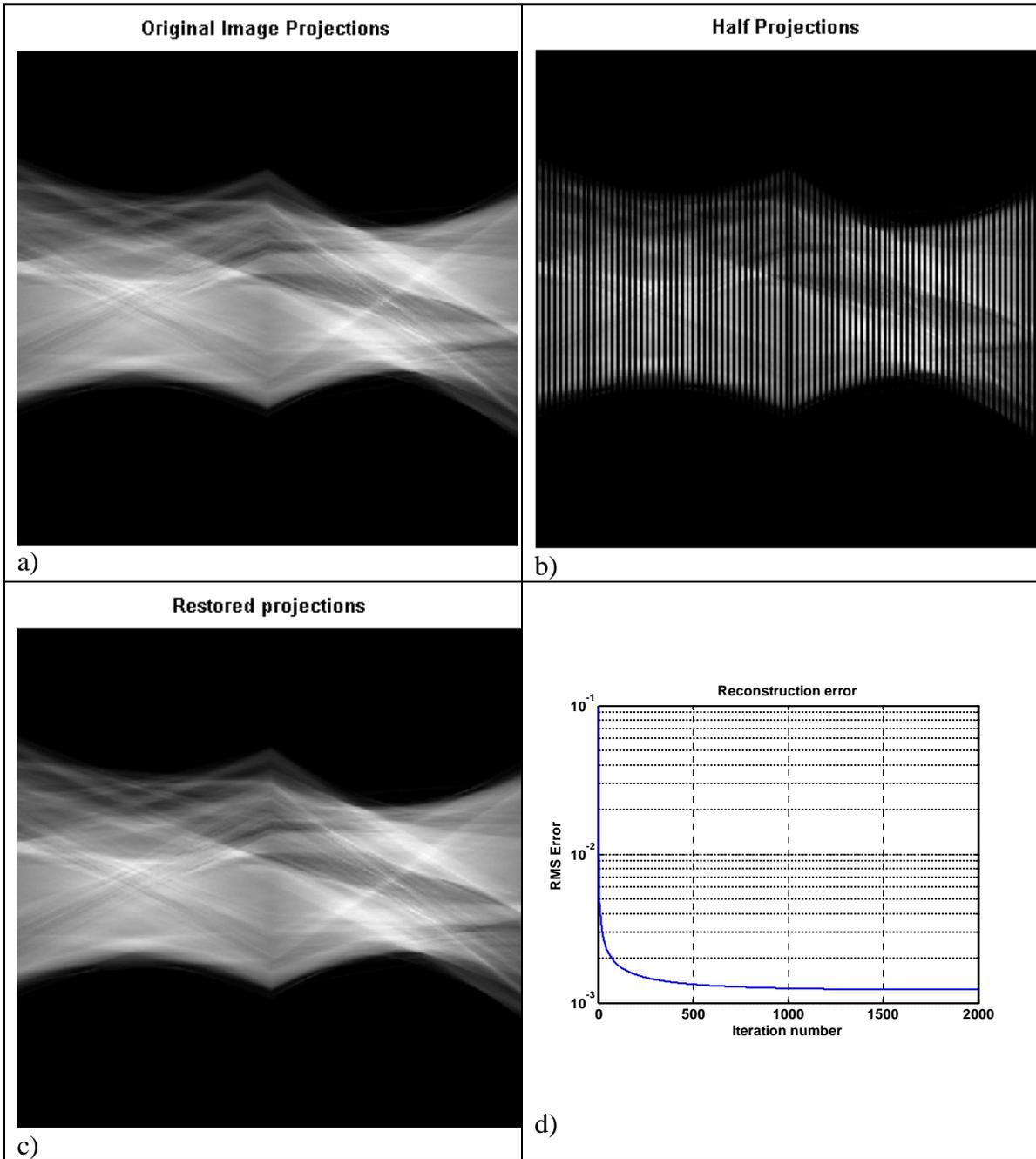

Fig. 11 - Recovery of missing image projections: (a), original projections (sinogram) of the test image of Fig. 10, a), (b) sinogram with every second projection removed; b) sinogram recovered from (b) using the iterative interpolation algorithm and (c ) plot of standard deviation of image reconstruction error as a function of the iteration number.

# 7   Conclusion

The paper addresses the problem of reconstruction of discrete signals from their irregular samples and recovery of missing data. Considering that positions of available signal samples are always specified with certain accuracy that defines maximal number of signal samples sufficient for signal representation, we suggest a new approach to optimal recovery of discrete signals from irregularly sampled or sparse data based on the Discrete Sampling Theorem introduced in Sect. 2. The discrete sampling theorem refers to discrete signals band-limited in a domain of a certain transform and states that "*KofN* band-limited*"* discrete signals of *N* samples, which have only *K* ≤ *N* non-zero transform coefficients, can be precisely recovered from their *K* sparse samples provided positions of the available samples satisfy certain limitations depended on the transform. This theorem provides also a tool for optimal, in terms of root mean squared error, approximation of arbitrary discrete signals specified by their sparse samples with "*KofN*- band-limited" signals, provided appropriate selection of the signal representation transform.

Two algorithms for discrete sampling theorem based signal reconstruction are considered, direct matrix inversion and Gershbrg-Papoulis iterative type iterative algorithm.

Properties of different transforms, such as Discrete Fourier, Discrete Cosine, Haar, Walsh and wavelet transforms, relevant to application of the Discrete Sampling Theorem are discussed and, in particular, it is shown that precise reconstruction of one-dimenstional "*KofN*-DFT band-limited" and "*KofN*-DCT band-limited" signals is always possible from sparse samples regardless of sample positions on the signal dense grid and that same holds for two-dimensional signals provided separable band-limitation conditions. For non-separable band limitation, such as limitation by a circle sector in DCT domain, experimental evidence is obtained that exact image recovery may not be possible for arbitrary placed samples and that redundant number of samples is required.

Applications of the discrete sampling theorem based approach to image recovery from sparse data are illustrated on examples of image super-resolution from multiple randomly sampled frames and image reconstruction from sparsely sampled projections. For the latter case, it is shown that, in applications, where object slices contain areas, which a priori are known to be empty, reconstruction of slice images from a given set of projections, is possible with super-resolution.